\documentclass[aps,showpacs]{revtex4}
\usepackage{color}
\usepackage{graphicx}
\usepackage{graphics}
\usepackage{epsfig}
\usepackage{subfigure}
\usepackage{latexsym}
\usepackage{amsfonts}
\usepackage[english]{babel}
\usepackage[latin1]{inputenc}
\usepackage[all]{xy}
\usepackage{amsmath}
\usepackage{amssymb}
\newcommand{\be}{\begin{equation}}
\newcommand{\ee}{\end{equation}}
\newcommand{\ben}{\begin{eqnarray}}
\newcommand{\een}{\end{eqnarray}}
\newcommand{\bes}{\begin{subequations}}
\newcommand{\ees}{\end{subequations}}

\newcommand{\ba}[1]{\begin{array}{#1}}
\newcommand{\ea}{\end{array}}
\newcommand{\bea}[1]{\begin{equation}\left\{\begin{array}{#1}}
\newcommand{\eea}{\end{array}\right.\end{equation}}

\begin{document}
\title{\textbf{On the semiclassical mass of ${\mathbb S}^2$-kinks}}

\author{A. Alonso-Izquierdo$^{a}$, M. A. Gonzalez Leon$^{a}$, J. Mateos Guilarte$^{b}$, and M. J.
Senosiain$^{c}$} \affiliation{{$^{a}$ Departamento de Matematica
Aplicada and IUFFyM, Universidad de Salamanca, SPAIN}
\\{$^{b}$Departamento de Fisica Fundamental and IUFFyM, Universidad de Salamanca, SPAIN}
\\{$^{c}$ Departamento de
Matematicas, Universidad de Salamanca, SPAIN}}

\begin{abstract}
One-loop mass shifts to the classical masses of stable kinks arising
in a massive non-linear ${\mathbb S}^2$-sigma model are computed.
Ultraviolet divergences are controlled using the heat kernel/zeta
function regularization method. A comparison between the results
achieved from exact and high-temperature asymptotic heat traces is
analyzed in depth.
\end{abstract}
 \maketitle

\section{Introduction}
In a seminal paper, Olive and Witten \cite{OW} linked extended
supersymmetric theories to BPS solitons by showing that the
classical mass of these stable lumps agreed exactly with the central
charge of the extended SUSY algebra. The subsequent issue concerning
BPS saturation at one-loop (rather than tree) level has proved to be
extremely subtle, prompting a remarkable amount of work over the
last twelve years. See, e.g., \cite{RvNW} and References quoted
therein to find an in-depth report on these developments.

A new actor entered the stage when in \cite{MRvNW} a Stony
Brook/Wien group computed the one-loop mass shift of the
supersymmetric ${\mathbb C}{\mathbb P}^1$-kink in a $N=(2,2)$
non-linear sigma model with twisted mass. Kinks of several types in
massive non-linear sigma models were, however, discovered earlier,
see \cite{AT1}, \cite{AT2}, \cite{ANNS}, \cite{D}. In Reference
\cite{AMAJ}, three of us found several families of non-topological
kinks in another non-linear sigma model: we chose ${\mathbb S}^2$ as
the target space and considered the case when the masses of the
pseudo-Nambu-Goldstone particles were different. The ${\mathbb
O}(2)$-symmetry of the equal-mass case is explicitly broken to
${\mathbb Z}_2\times{\mathbb Z}_2$ and the ${\mathbb S}{\mathbb
O}(2)$-families of topological kinks of the former system are
deformed to the four families of non-topological kinks arising in
the second system. The boundary of the moduli space of
non-topological kinks in the last model is formed by a pair of
topological kinks of different energy. The analysis of kink
stability in the massive non-linear ${\mathbb S}^2$-sigma model
performed in \cite{AMAJ1} allowed us to calculate the one-loop mass
shifts for the topological kinks by using the Cahill-Comtet-Glauber
formula \cite{CCG}. These authors showed that the one-loop mass
shift for static solitons can be read from the eigenvalues of the
bound states of the kink second-order fluctuation operator and the
threshold to the continuous spectrum when this operator  is a
transparent Schr$\ddot{\rm o}$dinger operator of the P$\ddot{\rm
o}$sch-Teller type. This is the case of the topological kinks of the
massive non-linear ${\mathbb S}^2$-sigma model when a parallel frame
to the kink orbits is chosen to refer to the fluctuations.

The aim of this paper is to offer another route for computing the
one-loop kink mass shift in order to unveil some of the intricacies
hidden in this subtle problem. We shall follow the method developed
in References \cite{AMAJW} and \cite{AMAJW1} based on heat
kernel/zeta function regularization of ultraviolet divergences. See
also the lectures \cite{AMAJWMJM}, where full details can be found.
Because the spectrum of small kink fluctuations in our system can be
identified analytically, we are able to give the exact answer for
the mass shifts. We shall also show, however, how to reach
approximately the same result using the coefficients of the heat
kernel asymptotic expansion. The interest of this calculation is
that a formula belonging to the class of formulas shown in
\cite{AMJW} will be derived. The importance of this type of formula
lies in the fact that it can be applied to obtain the one-loop mass
shifts of topological defects even when the spectrum of the
second-order fluctuation operator is not known; for instance, in the
case of two-component topological kinks: see \cite{AMAJW},
\cite{AMAJW1}. Similar formulas work even for Abelian gauge theories
in (2+1)-dimensions and thus the mass shifts of self-dual
Nielsen-Olesen vortices and semi-local strings can be calculated
approximately, see \cite{AMJW1}, \cite{AMJW2}, and \cite{AMJW3}.

To end this brief Introduction we simply mention that interesting
calculations have recently  appeared addressing one-loop kink mass
corrections and kink melting at finite temperatures in the
sine-Gordon, ${\mathbb CP}^1$, and $\lambda\phi^4$ models in a
purely bosonic setting, see \cite{RSvN}.

The organization of the paper is as follows: In Section \S.II, we
introduce the model and explain our conventions. In Section \S.III,
the perturbative sector as well as the mass renormalization
procedure are discussed. Section \S.IV is devoted to the analysis of
the stable topological kinks in this system. The second-order kink
fluctuation operator is obtained, placing special emphasis on its
geometric properties. In Section \S.V, the one-loop mass shift is
computed using the heat kernel/zeta function regularization method.
Section \S.VI offers a comparison of the exact result obtained in
\S.V with the approximation reached from the high-temperature
asymptotic expansion. Finally, a summary and outlook are offered
whereas two Appendices containing some technical material are
included.

\section{The (1+1)-dimensional massive non-linear ${\mathbb S}^2$-sigma model}
The action governing the dynamics of the non-linear ${\mathbb
S}^2$-sigma model and the constraint on the scalar fields are:
\begin{equation}
S[\phi_1,\phi_2,\phi_3]=\int \, dtdx \,
\left\{\frac{1}{2}g^{\mu\nu}\sum_{a=1}^3\frac{\partial\phi_a}{\partial
x^\mu}\frac{\partial\phi_a}{\partial x^\nu}\right\} \qquad , \qquad
\phi_1^2+\phi_2^2+\phi_3^2=R^2  \hspace{0.5cm} .
\end{equation}
The scalar fields are thus maps, $\phi_a(t,x)\in {\rm Maps}({\mathbb
R}^{1,1},{\mathbb S}^2)$, $a=1,2,3$, from the $(1+1)$-dimensional
Minkowski space-time to a ${\mathbb S}^2$-sphere of radius $R$,
which is the target manifold of this non-linear sigma model. Our
conventions for ${\mathbb R}^{1,1}$ are as follows: $x^\mu\in
{\mathbb R}^{1,1}$, $\mu=0,1$ with $x^0=t, x^1=x$ and
$g^{\mu\nu}={\rm diag}(1,-1)$. Then $x^\mu \cdot
x_\mu=g^{\mu\nu}x_\mu x_\nu = t^2-x^2$ and
\[
\frac{\partial}{\partial x_\mu}\left(\frac{\partial}{\partial
x^\mu}\right)=g^{\mu\nu}\frac{\partial^2}{\partial x^\mu\partial
x^\nu}=\Box=\frac{\partial^2}{\partial
t^2}-\frac{\partial^2}{\partial x^2} \qquad .
\]
The infrared asymptotics forbids massless particles in
$(1+1)$-dimensional scalar field theories, see \cite{Col}. We
therefore include the simplest potential energy density that would
be generated by quantum fluctuations {\footnote{ Without loss of
generality, we choose the parameters such that:
$\alpha_1^2\geq\alpha_2^2
> \alpha_3^2\geq 0$.}}:
\[
V(\phi_1,\phi_2,\phi_3)=\frac{1}{2} \left( \alpha_1^2 \,
\phi_1^2+\alpha_2^2 \, \phi_2^2+\alpha_3^2 \, \phi_3^2\right) \qquad
.
\]
\begin{enumerate}

\item Solving $\phi_3$ in favor of $\phi_1$ and $\phi_2$, \,\,  $
{\rm sg}(\phi_3)\phi_3=\sqrt{R^2-\phi_1^2-\phi_2^2}$, we find:
\[
S={1\over 2}\int \, dtdx \, \left\{\frac{\partial\phi_1}{\partial
x_\mu}\cdot\frac{\partial\phi_1}{\partial
x^\mu}+\frac{\partial\phi_2}{\partial
x_\mu}\cdot\frac{\partial\phi_2}{\partial
x^\mu}+\frac{(\phi_1\partial_\mu\phi_1+\phi_2\partial_\mu\phi_2
)}{\sqrt{R^2-\phi_1^2-\phi_2^2}}\cdot\frac{(\phi_1\partial^\mu\phi_1+\phi_2\partial^\mu\phi_2
)}{\sqrt{R^2-\phi_1^2-\phi_2^2}}-V_{{\mathbb
S}^2}[\phi_1,\phi_2]\right\} \, \, ,
\]
where
\[
V_{{\mathbb S}^2}(\phi_1,\phi_2)=\frac{1}{2} \left(
(\alpha_1^2-\alpha_3^2) \, \phi_1^2+(\alpha_2^2-\alpha_3^2) \,
\phi_2^2+{\rm const.}
\right)\simeq\frac{\lambda^2}{2}\phi_1^2(t,x)+\frac{\gamma^2}{2}\phi_2^2(t,x)\quad
,
\]
with $\lambda^2=(\alpha_1^2-\alpha_3^2)$,
$\gamma^2=(\alpha_2^2-\alpha_3^2)$, $\lambda^2\geq \gamma^2$. The
masses of the pseudo-Nambu-Goldstone bosons are respectively
$\lambda$ and $\gamma$.

\item Interactions, however, come from the geometry:
\begin{eqnarray*}
&&\frac{(\phi_1\partial_\mu\phi_1+\phi_2\partial_\mu\phi_2
)}{\sqrt{R^2-\phi_1^2-\phi_2^2}}\cdot\frac{(\phi_1\partial^\mu\phi_1+\phi_2\partial^\mu\phi_2
)}{\sqrt{R^2-\phi_1^2-\phi_2^2}}\simeq \\
&\simeq& {1\over R^2}\left(1+{1\over R^2}(\phi_1^2+\phi_2^2)+{1\over
R^4}(\phi_1^2+\phi_2)^2+\cdots\right)\cdot\left(
\phi_1\frac{\partial\phi_1}{\partial
x^\mu}+\phi_2\frac{\partial\phi_2}{\partial x^\mu}\right)\left(
\phi_1\frac{\partial\phi_1}{\partial
x_\mu}+\phi_2\frac{\partial\phi_2}{\partial x_\mu}\right)
\end{eqnarray*}
and ${1\over R^2}$ is a non-dimensional coupling constant.

\end{enumerate}
In the unit natural system, $\hbar=c=1$, the dimensions of fields,
masses and coupling constants are respectively: $ [\phi_a]=1=[R]$,
$[\gamma]=M=[\lambda]$. We define non dimensional space-time
coordinates and masses:
\[
x^\mu \quad \longrightarrow \quad \frac{x^\mu}{\lambda} \qquad
\qquad , \qquad \qquad
\sigma^2=\frac{\alpha_2^2-\alpha_3^2}{\alpha_1^2-\alpha_3^2}=\frac{\gamma^2}{\lambda^2}
\qquad ,\qquad 0 < \sigma^2 \leq 1
\]
to write the action and the energy in terms of them:
\begin{equation}
S={1\over 2}\int \, dtdx \,  \left\{\frac{\partial\phi_1}{\partial
x_\mu}\cdot\frac{\partial\phi_1}{\partial
x^\mu}+\frac{\partial\phi_2}{\partial
x_\mu}\cdot\frac{\partial\phi_2}{\partial x^\mu} +
\frac{(\phi_1\partial_\mu\phi_1+\phi_2\partial_\mu\phi_2)}{\sqrt{R^2-\phi_1^2-\phi_2^2}}
\cdot\frac{(\phi_1\partial^\mu\phi_1+\phi_2\partial^\mu\phi_2)}{\sqrt{R^2-\phi_1^2-\phi_2^2}}
-\phi_1^2(t,x)-\sigma^2\phi_2^2(t,x)\right\} \label{eq:act1}
\end{equation}
\begin{eqnarray*}
E={\lambda\over 2}\int \, &dx& \,
\left\{\left(\frac{\partial\phi_1}{\partial
t}\right)^2+\left(\frac{\partial\phi_2}{\partial t}\right)^2 +
\frac{(\phi_1\partial_t\phi_1+\phi_2\partial_t\phi_2)^2}{R^2-\phi_1^2-\phi_2^2}+
\left(\frac{\partial\phi_1}{\partial
x}\right)^2+\left(\frac{\partial\phi_2}{\partial x}\right)^2
+\frac{(\phi_1\partial_x\phi_1+\phi_2\partial_x\phi_2)^2}{R^2-\phi_1^2-\phi_2^2} \right. \\
&+& \left. \phi_1^2(t,x)+\sigma^2\cdot\phi_2^2(t,x)\right\}  \qquad
.
\end{eqnarray*}

There are two homogeneous minima of the action or vacua of our
model: $ \phi_1^{V^\pm}=\phi_2^{V^\pm}=0 \,\,  , \,\,
\phi_3^{V^\pm}= \pm R$, the North and South Poles. Choice of one of
the poles to quantize the system spontaneously breaks the ${\mathbb
Z}_2\times{\mathbb Z}_2\times{\mathbb Z}_2$ symmetry of the action
(\ref{eq:act1}), $\phi_a \rightarrow (-1)^{\delta_{ab}}\phi_b \, \,
, \, \, a,b=1,2,3$, to: ${\mathbb Z}_2\times{\mathbb Z}_2 \, \, \, ,
\, \, \, \phi_\alpha \rightarrow
(-1)^{\delta_{\alpha\beta}}\phi_\beta  \, ,  \,  \alpha \, ,
\,\beta=1,2$. Therefore, the configuration space ${\cal
C}=\left\{{\rm Maps}({\mathbb R},{\mathbb S}^2)/E<+\infty \right\}$
is the union of four disconnected sectors ${\cal C}={\cal C}_{\rm
NN}\bigcup {\cal C}_{\rm SS} \bigcup {\cal C}_{\rm NS} \bigcup {\cal
C}_{\rm SN}$ labeled by the vacua reached by each configuration at
the two disconnected components of the boundary of the real line:
$x=\pm\infty $.

\section{Mass renormalization}
The field equations
\begin{eqnarray*}
\Box \phi_1+\partial_\mu
\left[\frac{\phi_1\sum_{\alpha=1}^2\phi_\alpha\partial^\mu\phi_\alpha}{R^2-
\sum_{\alpha=1}^2\phi_\alpha\phi_\alpha}\right]
+\phi_1\left[\frac{\sum_{\alpha=1}^2\phi_\alpha\partial_\mu\phi_\alpha
\sum_{\beta=1}^2\phi_\beta\partial^\mu\phi_\beta}{(R^2-\sum_{\alpha=1}^2\phi_\alpha\phi_\alpha)^{3\over
2}}+1\right]&=&0 \\ \Box \phi_2+\partial_\mu
\left[\frac{\phi_2\sum_{\alpha=1}^2\phi_\alpha\partial^\mu\phi_\alpha}{R^2-
\sum_{\alpha=1}^2\phi_\alpha\phi_\alpha}\right]
+\phi_2\left[\frac{\sum_{\alpha=1}^2\phi_\alpha\partial_\mu\phi_\alpha
\sum_{\beta=1}^2\phi_\beta\partial^\mu\phi_\beta}{(R^2-\sum_{\alpha=1}^2\phi_\alpha\phi_\alpha)^{3\over
2}}+\sigma^2\right]&=&0
\end{eqnarray*}
become linear for small fluctuations,
$G_\alpha(x^\mu)=\phi_\alpha^{V^\pm}+\delta G_\alpha(x^\mu)$, around
the vacuum:
\begin{equation}
\Box \delta G_1(t,x)+\delta G_1(t,x)={\cal O}(\delta G_\alpha\delta
G_\beta) \qquad \quad , \qquad \quad \Box \delta
G_2(t,x)+\sigma^2\delta G_2(t,x)={\cal O}(\delta G_\alpha\delta
G_\beta) \qquad . \label{lfeq}
\end{equation}
We shall need the Feynman rules only for the four-valent vertices.
Besides the two propagators for the (pseudo) Nambu-Goldstone bosons
there are three vertices with four external legs. The derivatives
appearing in the interactions induce dependence on the momenta in
the weights. This also affects the sign and the combinatorial
factors. Naturally, there are many more vertices in this model, but
we list only the vertices that contribute to the self-energy of the
Nambu-Goldstone bosons up to one-loop order.

\begin{table}[h]
\begin{center}
\caption{Propagators}
\begin{tabular}{lccc} \\ \hline
\textit{Particle} & \textit{Field} & \textit{Propagator} & \textit{Diagram} \\
\hline \\ Nambu-Goldstone & $G_1(x^\mu)$ &
$\displaystyle\frac{i}{k_0^2-k^2-1+i\varepsilon}$ &
\parbox{3cm}{\epsfig{file=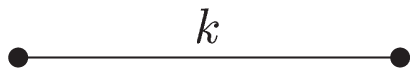,width=3cm}}
\\[0.5cm]
Nambu-Goldstone & $G_2(x^\mu)$ &
$\displaystyle\frac{i}{k_0^2-k^2-\sigma^2+i\varepsilon}$ &
\parbox{3cm}{\epsfig{file=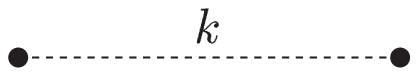,width=3cm}} \\[0.5cm] \hline
\end{tabular}
\end{center}
\end{table}

\begin{table}[hbt]
\begin{center}
\caption{Fourth-order vertices }
\begin{tabular}{clclcl} \\ \hline
\textit{Vertex} & \textit{Weight} & \textit{Vertex} & \textit{Weight} & \textit{Vertex} & \textit{Weight} \\
\hline \\
\parbox{2.8cm}{\epsfig{file=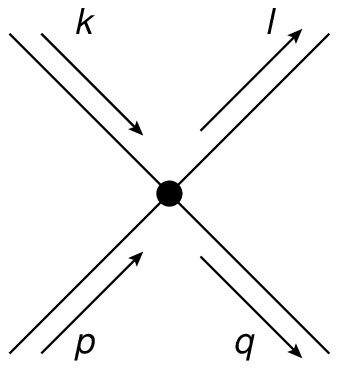,width=2.2cm}} &
$\displaystyle 2i\frac{k_\mu p^\mu}{R^2}$  &
\parbox{2.8cm}{\epsfig{file=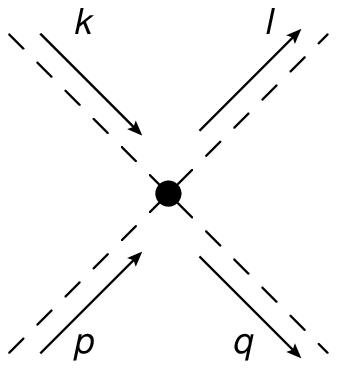,width=2.2cm}} &
$\displaystyle 2i\frac{k_\mu p^\mu}{R^2}$ &
\parbox{2.8cm}{\epsfig{file=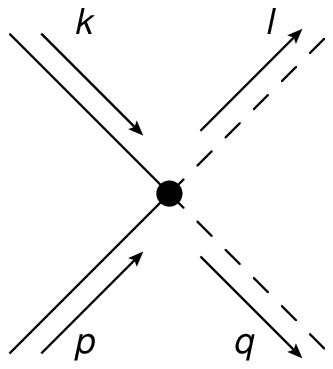,width=2.2cm}} &
$\displaystyle 2i\frac{k_\mu p^\mu}{R^2}$
\\[0.7cm] \hline
\end{tabular}
\end{center}
\end{table}

\subsection{Plane waves and vacuum energy}

The general solution of the linearized field equations (\ref{lfeq})
governing the small fluctuations of the Nambu-Goldstone fields is:
\begin{eqnarray*}
\delta G_1(x_0,x)&=&\frac{1}{2}\cdot\sqrt{{1\over
l}}\sum_k{1\over\sqrt{2\omega_1(k)}}
\left\{a_1(k)e^{-ik_0x_0+ikx}+a_1^*(k)e^{ik_0x_0-ikx}\right\}
\\\delta
G_2(x_0,x)&=&\frac{1}{2}\cdot\sqrt{\frac{1}{l}}\sum_q{1\over\sqrt{2\omega_2(q)}}
\left\{a_2(q)e^{iq_0x_0-iqx}+a_2^*(q)e^{-iq_0x_0+iqx}\right\} \qquad
,
\end{eqnarray*}
where $k_0=\omega_1(k)=\sqrt{k^2+1}$,
$q_0=\omega_2(q)=\sqrt{q^2+\sigma^2}$, and the dispersion relations
$k_0^2-k^2-1=0$, $q_0^2-q^2-\sigma^2=0$ hold:
\[
K_0\left(\begin{array}{c} e^{ikx} \\ 0
\end{array}\right)=\omega_1^2(k)\left(\begin{array}{c} e^{ikx} \\ 0
\end{array}\right)\qquad , \qquad K_0\left(\begin{array}{c} 0 \\e^{iqx}
\end{array}\right)=\omega_2^2(q)\left(\begin{array}{c} 0 \\ e^{iqx}
\end{array}\right)
\]
\[
K_0=\left(\begin{array}{cc} K_{011} & 0 \\ 0 & K_{022}\end{array}
\right)=\left(\begin{array}{cc} -{d^2\over dx^2}+1 & 0
\\ 0 & -{d^2\over dx^2}+\sigma^2\end{array} \right) \qquad .
\]
We have chosen a normalization interval of non-dimensional \lq\lq
length" $l=\lambda L$, $I=[-\frac{l}{2},\frac{l}{2}]$, and we impose
PBC on the plane waves so that: $k\cdot l=2\pi n_1$ , $q\cdot l=2\pi
n_2$ with $n_1,n_2\in{\mathbb Z}$. Thus, $K_0$ acts on
$L^2=\bigoplus_{\alpha=1}^2 L_\alpha^2({\mathbb S}^1)$, and its
spectral density at the $l\to\infty$ limit is: $
\rho_{K_0}(k)=\left(\begin{array}{cc} \frac{dn_1}{dk} & 0
\\ 0 & \frac{dn_2}{dq} \end{array}\right)={l\over 2\pi} \left(\begin{array}{cc} 1 & 0
\\ 0 & 1 \end{array}\right)$.

From the classical free (quadratic) Hamiltonian
\begin{eqnarray*}
H^{(2)}&=&{\lambda\over 2}\int \, dx \,
\left\{\left(\frac{\partial\delta G_1}{\partial
x_0}\cdot\frac{\partial\delta G_1}{\partial
x_0}+\frac{\partial\delta G_1}{\partial x}\cdot\frac{\partial\delta
G_1}{\partial x}\right)+\left(\frac{\partial\delta G_2}{\partial
x_0}\cdot\frac{\partial\delta G_2}{\partial
x_0}+\frac{\partial\delta G_2}{\partial x}\cdot\frac{\partial\delta
G_2}{\partial x}\right)\right.\\&+&\left. \delta G_1\cdot\delta
G_1+\sigma^2\delta G_2\cdot\delta G_2\right\}=\sum_k \,
\sum_{\alpha=1}^2 \, {\lambda\over 2}\left[
\omega_\alpha(k)(a_\alpha^*(k)a_\alpha(k)+a_\alpha(k)a_\alpha^*(k))\right]\qquad
,
\end{eqnarray*}
one passes via canonical quantization to the quantum free
Hamiltonian:
\[
[\hat{a}_\alpha(k),\hat{a}^\dagger_\beta(q)]=\delta_{\alpha\beta}\delta_{kq}
\qquad , \qquad \hat{H}^{(2)}_0=\sum_k \,  \lambda\left[
\omega_1(k)\left(\hat{a}_1^\dagger(k)\hat{a}_1(k)+{1\over
2}\right)+\omega_2(k)\left(\hat{a}_2^\dagger(k)\hat{a}_2(k)+{1\over
2}\right)\right] \qquad .
\]
The vacuum energy is:
\[
\hat{a}_\alpha(k)|0;V\rangle =0 \, \, , \, \, \forall k \, , \,
\forall \alpha \qquad , \qquad \Delta E_0=\langle
0;V|\hat{H}^{(2)}_0|0;V\rangle ={\lambda\over
2}\sum_k\omega_1(k)+{\lambda\over 2}\sum_k\omega_2(k)={\lambda\over
2}{\rm Tr}_{L^2} K_0^{{1\over 2}}
\]

\subsection{One-loop mass renormalization counter-terms}

There are four ultraviolet divergent graphs in one-loop order of the
$\hbar$-expansion contributing to the $G_1(x^\mu)$ and $G_2(x^\mu)$
Nambu-Goldstone bosons self-energies:
\begin{itemize}
\item Self-energy of $G_2$
\begin{eqnarray*}
&& {2i\over R^2}\cdot I(1)+{2i\over R^2}\cdot
I(\sigma^2)=\hspace{1cm}\parbox{3cm}{\epsfig{file=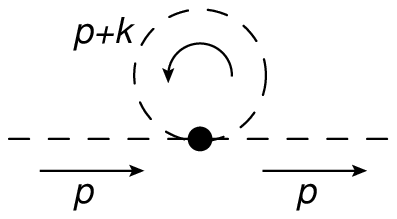,width=3cm}}+
\parbox{4cm}{\epsfig{file=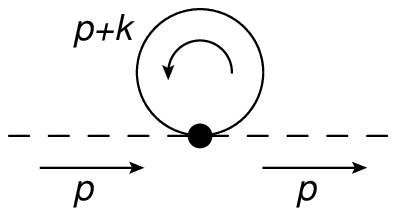,width=3cm}}\\&=&{2i\over R^2}\cdot
\int \,\frac{d^2k}{(2\pi)^2}\, \cdot \frac{i(p_\mu+k_\mu)p^\mu
}{(p_\mu+k_\mu) k^\mu -1+i\varepsilon}+{2i\over  R^2}\cdot \int
\,\frac{d^2k}{(2\pi)^2}\, \cdot
\frac{i(p_\mu+k_\mu)p^\mu}{(p_\mu+k_\mu)
k^\mu-\sigma^2+i\varepsilon}\\&=&{2i\over R^2}\cdot \int
\,\frac{d^2k}{(2\pi)^2}\, \cdot \frac{i }{k_\mu k^\mu
-1+i\varepsilon}+{2i\over  R^2}\cdot \int \,\frac{d^2k}{(2\pi)^2}\,
\cdot \frac{i}{k_\mu k^\mu-\sigma^2+i\varepsilon} \qquad ,
\end{eqnarray*}
\item Self-energy of $G_1$
\begin{eqnarray*} && {2i\sigma^2\over R^2}\cdot I(1)+{2i\sigma^2\over R^2}\cdot
I(\sigma^2)=\hspace{1cm}\parbox{3cm}{\epsfig{file=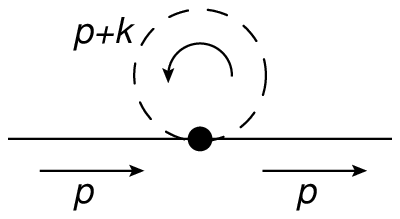,width=3cm}}
+
\parbox{3cm}{\epsfig{file=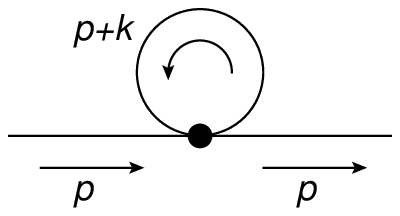,width=3cm}}\\&=&{2i\sigma^2\over R^2}\cdot \int \,\frac{dk}{4\pi}\,
\cdot \frac{1}{\sqrt{k^2+1}}+{2i\sigma^2\over R^2}\cdot \int
\,\frac{dk}{4\pi}\, \cdot \frac{1}{\sqrt{k^2+\sigma^2}}\qquad ,
\end{eqnarray*}
where we have computed the $k_0$ integrations by using the residue
theorem. We only show this step explicitly in the computation of the
self-energy of $\delta G_1$ because it suffices to point out how to
regularize these divergent integrals by means of spectral zeta
functions. The regularization just mentioned will be performed later
in Section \S.V D.

The $p_\mu p^\mu$ factor becomes constant when the momentum is put
\lq\lq on shell" in the external legs, $p_\mu p^\mu=1$, $p_\mu
p^\mu=\sigma^2$. This process gives us the mass renormalization
counter-terms. The Lagrangian density of counter-terms shown in
Table III must be added to cancel the above divergences exactly. We
also show the vertices generated at one-loop level.
\end{itemize}
\begin{table}[hbt]
\caption{One-loop counter-terms}
\begin{tabular}{ccc} \\ \hline
\textit{Diagram} & \hspace{0.3cm} & \textit{Weight} \\
\hline
\parbox{4cm}{\epsfig{file=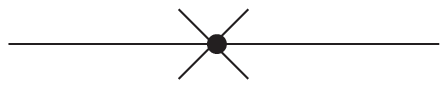,width=4cm}} & &
$\displaystyle -\frac{2i}{R^2}(I(1)+I(\sigma^2))$ \\
 \parbox{4cm}{\epsfig{file=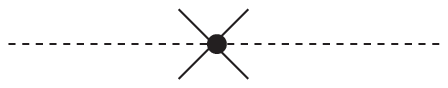,width=4cm}} & &
$\displaystyle  -\frac{2i}{R^2}(I(1)+I(\sigma^2))$\\[0.5cm] \hline \\[0.25cm]
\end{tabular}\hspace{0.7cm} {\normalsize $\displaystyle {\cal L}_{C.T.}=-{1\over R^2}\cdot\left[ I(1)+I(\sigma^2)\right]
\left(\phi_1^2(x^\mu)+\sigma^2\phi_2^2(x^\mu)\right)$}
\end{table}

\section{Isothermal coordinates and topological kinks}

In this Section we shall use the isothermal coordinates in the chart
${\mathbb S}^2-\{(0,0,-R)\}$ obtained via stereographic projection
from the South Pole:
\begin{equation}
\chi^1=\frac{\phi_1}{1+\frac{\phi_3}{R}}=\frac{R \phi_1}{R+{\rm
sg}(\phi_3)\sqrt{R^2-\phi_1^2-\phi_2^2}} \qquad , \qquad
\chi^2=\frac{\phi_2}{1+\frac{\phi_3}{R}}=\frac{R \phi_2}{R+{\rm
sg}(\phi_3)\sqrt{R^2-\phi_1^2-\phi_2^2}} \qquad . \label{iso}
\end{equation}
The geometric data of the sphere in this coordinate system are:
\begin{eqnarray*}
ds^2&=&\frac{4 R^4}{(R^2+\chi^1\chi^1+\chi^2\chi^2)^2} \qquad ,
\qquad g_{11}(\chi^1,\chi^2)=g_{22}(\chi^1,\chi^2)=\frac{4
R^4}{(R^2+\chi^1\chi^1+\chi^2\chi^2)^2} \\
\Gamma^1_{11}(\chi^1,\chi^2)&=&-\Gamma^1_{22}(\chi^1,\chi^2)=
\Gamma^2_{12}(\chi^1,\chi^2)=\Gamma^2_{21}(\chi^1,\chi^2)=\frac{-2
\chi^1}{R^2+\chi^1\chi^1+\chi^2\chi^2}\\
\Gamma^2_{22}(\chi^1,\chi^2)&=&-\Gamma^2_{11}(\chi^1,\chi^2)=\Gamma^1_{12}(\chi^1,\chi^2)
=\Gamma^1_{21}(\chi^1,\chi^2)= \frac{-2
\chi^2}{R^2+\chi^1\chi^1+\chi^2\chi^2}
\\
R^1_{122}(\chi^1,\chi^2)&=&-R^1_{212}(\chi^1,\chi^2)=-R^2_{121}(\chi^1,\chi^2)=
R^2_{211}(\chi^1,\chi^2)= \frac{-4
R^4}{(R^2+\chi^1\chi^1+\chi^2\chi^2)^2} \quad .
\end{eqnarray*}
The kinetic and potential energy densities read:
\[
T(\chi^1,\chi^2)=\frac{2 R^4}{(R^2+\chi^1\chi^1+\chi^2\chi^2)^2}
\cdot\left(\partial_t\chi^1\partial_t\chi^1
+\partial_t\chi^2\partial_t\chi^2 \right)
\]
\[
V(\chi^1,\chi^2)=\frac{2 R^4}{(R^2+\chi^1\chi^1+\chi^2\chi^2)^2}
\cdot\left(\partial_x\chi^1\partial_x\chi^1
+\partial_x\chi^2\partial_x\chi^2
+\chi^1\chi^1+\sigma^2\chi^2\chi^2\right) \qquad .
\]
From the action $S=\int \, d^2x \, \left[T-V\right]$ one derives the
field equations:
\begin{equation}
\Box \chi^i+\Gamma^i_{jk}\partial_\mu\chi^j\partial^\mu\chi^k
+\delta^i_1\chi^1+\sigma^2\delta^i_2\chi^2-2(\delta^i_1\chi^1+\delta^i_2\chi^2)
\frac{\chi^1\chi^1+\sigma^2\chi^2\chi^2}{R^2+\chi^1\chi^1+\chi^2\chi^2}=0
\label{feq} \quad ,
\end{equation}
which for static configurations reduce to:
\begin{equation}
-\frac{d^2\chi^i}{dx^2}-\Gamma^i_{jk}\frac{d\chi^j}{dx}\frac{\chi^k}{dx}
+\delta^i_1\chi^1+\sigma^2\delta^i_2\chi^2-2(\delta^i_1\chi^1+\delta^i_2\chi^2)
\frac{\chi^1\chi^1+\sigma^2\chi^2\chi^2}{R^2+\chi^1\chi^1+\chi^2\chi^2}=0
\label{sfeq} \qquad .
\end{equation}

\subsection{Topological $K$ kinks}
We try the $\chi^1=0$ orbit in (\ref{sfeq}) and reduce this ODE
system to the single ODE:
\begin{equation}
\frac{d^2\chi^2}{dx^2}-\frac{2\chi^2}{R^2+\chi^2\chi^2}\frac{d\chi^2}{dx}\frac{d\chi^2}{dx}=\sigma^2\chi^2
\left(1-2\frac{\chi^2\chi^2}{R^2+\chi^2\chi^2}\right) \label{t1sfeq}
\qquad .
\end{equation}
\begin{equation}
\chi^1_{{\rm K}}(x)=0  \qquad \quad , \quad \qquad \chi^2_{{\rm
K}}(x)=\pm R e^{\pm\sigma (x-x_0)} \qquad , \label{t1kp}
\end{equation}
are solutions of (\ref{t1sfeq}) of finite energy:
\begin{equation}
E[K]=\lambda \int_{-\infty}^\infty \, dx \, \frac{R^2\sigma^2}{{\rm
cosh}^2(\sigma(x-x_0))}=2\lambda R^2\sigma \label{t1ken} \qquad .
\end{equation}
In (\ref{t1kp}), $x_0$ is an integration constant that sets the kink
center. The kink field components in the original coordinates
\[
\phi_1^{K}(x)=0 \qquad , \qquad \phi_2^{K}(x)=\frac{R}{{\rm
cosh}[\sigma(x-x_0)]} \qquad , \qquad \phi_3^{K}(x)=\pm R\,{\rm
tanh}[\sigma(x-x_0)]
\]
are either kink-shaped, $\phi_3^{{\rm K}}$, or bell-shaped,
$\phi_2^{{\rm K}}$. It is clear that the four solutions (\ref{t1kp})
belong to the topological sectors ${\cal C}_{\rm NS}$ or ${\cal
C}_{\rm SN}$ of the configuration space. Lorentz invariance tells us
that
\begin{equation}
\chi^1_{{\rm K}}(x)=0  \qquad \quad , \quad \qquad \chi^2_{{\rm
K}}(x)=\pm R \, \, {\rm exp}[\pm\sigma
(\frac{x-vt}{\sqrt{1-v^2}}-x_0)] \label{t1kplb}
\end{equation}
are solitary wave solutions of the full field equations (\ref{feq}).

\subsection{Second-order fluctuation operator}

Let us consider small kink fluctuations:
\[
\chi(x)=\chi_K(x)+\eta(x) \qquad \qquad , \qquad \qquad
\eta(x)=(\eta^1(x),\eta^2(x)) \qquad .
\]
Here, $\chi_K(x)=(\chi_K^1(x),\chi_K^2(x))$ is the kink solution and
$\eta
(x)=\eta^1(x)\frac{\partial}{\partial\chi^1}+\eta^2(x)\frac{\partial}{\partial\chi^2}\in
\Gamma(T{\mathbb S}^2)$ are vector fields along the kink orbit -
expressed in the orthonormal basis $\langle
\frac{\partial}{\partial\chi^i},\frac{\partial}{\partial\chi^j}\rangle=\delta^{ij}$
of $T{\mathbb S}^2$ - giving the small fluctuations on the kink.
From the tangent vector field to the orbit
$\chi^\prime_K(x)=\frac{d\chi^1_K}{dx}\frac{\partial}{\partial\chi^1}+\frac{d\chi^2_K}{dx}
\frac{\partial}{\partial\chi^2}$, the covariant derivative, and the
curvature tensor
\[
\nabla_{\chi_K^\prime} \eta(x)=\left(\eta^{\prime
i}(x)+\Gamma^i_{jk}(\chi_K)\eta^j(x)\chi_K^{\prime
k}(x)\right)\frac{\partial}{\partial\chi^i} \qquad , \qquad
R(\chi^\prime_K,\eta)\chi^\prime_K=\chi^{\prime
i}_K(x)\eta^j(x)\chi^{\prime
k}(x)R^l_{ijk}(\chi_K)\frac{\partial}{\partial\chi^l}\, \, \, \, \,
,
\]
we obtain the geodesic deviation operator:
\[
\frac{D^2\eta}{dx^2}(x)=\nabla_{\chi_K^\prime}\nabla_{\chi_K^\prime}\eta(x)
\qquad , \qquad \frac{D^2\eta}{dx^2}(x)+
R(\chi^\prime_K,\eta)\chi^\prime_K(x) \qquad .
\]
We also need the Hessian of the \lq\lq mechanical" potential
$U(\chi^1,\chi^2)=-V(\chi^1,\chi^2)$
\[
\nabla_\eta {\rm grad}U(x)=\eta^i(x)\left(\frac{\partial^2
U}{\partial\chi^i\partial\chi^j}(\chi_K)-\Gamma^k_{ij}(\chi_K)\frac{\partial
U}{\partial\chi^k}(\chi_K)\right)g^{jl}\frac{\partial}{\partial\chi^l}
\]
The second-order fluctuation operator around the kink $\chi_K$ is:
\begin{equation}
\Delta(K)\eta(x)=-\left[
\frac{D^2\eta}{dx^2}(x)+R(\chi^\prime_K,\eta)\chi^\prime_K
+\nabla_\eta {\rm grad}U(x)\right] \qquad . \label{sodk1}
\end{equation}

\subsection{Small fluctuations on $K$
kinks}

Application to the $K$ kink
$\chi_{K}(x)=(\chi_{K}^1(x)=0,\chi_{K}^2(x)=R\, e^{-\sigma x})$
gives:
\begin{eqnarray}
\Delta(K)\eta &=&\left[ -\left(\frac{d^2\eta^1}{dx^2}+2\sigma
(1-{\rm tanh}\sigma
x))\frac{d\eta^1}{dx}-\left(1-2\sigma^2+2\sigma^2{\rm tanh}\sigma
x\right)\eta^1\right)\frac{\partial}{\partial\chi^1}\right. \nonumber \\
&-&\left.\left( \frac{d^2\eta^2}{dx^2}+2\sigma (1-{\rm tanh}\sigma
x))\frac{d\eta^2}{dx}+\sigma^2\left(1-2{\rm tanh}\sigma
x\right)\eta^2\right)\frac{\partial}{\partial\chi^2}\right] \qquad .
\label{sodok1o}
\end{eqnarray}
The second-order fluctuation operator in the orthonormal frame is a
second-order differential operator that has first-order derivatives
both in the direction of the kink orbit,
$\frac{\partial}{\partial\chi^2}$, and the orthogonal direction to
the orbit $\frac{\partial}{\partial\chi^1}$.

Alternatively, we can use a parallel frame,
$\mu(x)=\mu^1(x)\frac{\partial}{\partial\chi^1}+\mu^2(x)
\frac{\partial}{\partial\chi^2}$, along the $K$ kink orbit:
\[
\frac{d\mu^i}{dx}+\Gamma^i_{jk}(\chi_K)\chi_K^{\prime j}\mu^k=0
\quad \equiv \quad \left\{
\begin{array}{c}\frac{d\mu^1}{dx}+\sigma(1-{\rm tanh})\mu^1(x)=0 \qquad
\Rightarrow \qquad \mu^1(x)=1+e^{-2\sigma x} \\
\\ \frac{d\mu^2}{dx}+\sigma(1-{\rm tanh})\mu^2(x)=0 \qquad \Rightarrow
\qquad \mu^2(x)=1+e^{-2\sigma x}\end{array}\right. \qquad .
\]
In this parallel frame the vectors of the basis
$\mu^i(x)\frac{\partial}{\partial\chi^i}$ point in the same
directions as $\frac{\partial}{\partial\chi^i}$ but their moduli
vary along the kink orbit:
\[
\langle
\mu^1(x)\frac{\partial}{\partial\chi^1},\mu^1(x)\frac{\partial}{\partial\chi^1}\rangle=
\langle
\mu^2(x)\frac{\partial}{\partial\chi^2},\mu^2(x)\frac{\partial}{\partial\chi^2}\rangle=(1+e^{-2\sigma
x})^2 \qquad .
\]
Writing the fluctuations in this frame,
$\eta(x)=\xi^1(x)\mu^1(x)\frac{\partial}{\partial\chi^1}+
\xi^2(x)\mu^2(x)\frac{\partial}{\partial\chi^2}$, we find:
\begin{equation}
\Delta(K)\eta=\mu^1(x)\left(
-\frac{d^2\xi^1}{dx^2}+(1-\frac{2\sigma^2}{{\rm cosh}^2\sigma
x})\xi^1\right)\, \frac{\partial}{\partial\xi^1} +\mu^2(x)\left(
-\frac{d^2\xi^2}{dx^2}+(\sigma^2-\frac{2\sigma^2}{{\rm cosh}^2\sigma
x})\xi^2\right)\, \frac{\partial}{\partial\chi^2} \label{sodok1p}
\qquad .
\end{equation}
In the parallel frame the second-order fluctuation operator is a
transparent (reflection coefficient equal to zero) P$\ddot{\rm
o}$sch-Teller Schr$\ddot{\rm o}$dinger operator both in the parallel
and orthogonal directions to the kink orbit.

This analysis is deceptively simple: acting respectively on
$\eta^1(x)=(1+e^{-2\sigma x})\xi^1(x)$ and $\eta^2(x)=(1+e^{-2\sigma
x})\xi^2(x)$ the terms with first-order derivatives in
(\ref{sodok1o}) disappear and $(1+e^{-2\sigma x})$ factors out,
leaving very well known Schr$\ddot{\rm o}$dinger operators acting
respectively on $\xi^1(x)$ and $\xi^2(x)$. The key point is that the
differential operators in (\ref{sodok1o}) and (\ref{sodok1p}) share
the eigenvalues although their eigenfunctions differ by the
$\mu^i(x)$ factors. The spectral functions associated are thus
identical and it seems wise to use the best known form. What we have
shown here is the geometrical meaning of the $\mu^i(x)$ factors:
they provide a parallel frame along the kink orbit.

\subsection{The spectrum of small kink fluctuations}
Changing from vector to matrix notation,
\[
\frac{\partial}{\partial\chi^1} \quad \longrightarrow \quad
\left(\begin{array}{c} 1 \\ 0 \end{array}\right) \quad \qquad ,
\quad \qquad \frac{\partial}{\partial\chi^2} \quad \longrightarrow
\quad \left(\begin{array}{c} 0 \\ 1 \end{array}\right) \qquad ,
\]
we now use the differential operators of formula (\ref{sodok1p}) to
write the linearized field equations satisfied by the small kink
fluctuations in the parallel frame:
\begin{eqnarray*}
\chi^1(t,x)=\chi^1_{{\rm K}}(x)+\mu^1(x)\delta K_1(t,x) \qquad &,&
\qquad \chi^2(t,x)=\chi^2_{{\rm K}}(x)+\mu^2(x)\delta K_2(t,x)\\
\frac{\partial^2\delta K_1}{\partial t^2}+K_{11}\delta K_1=0 \quad
\qquad &,& \quad \qquad \frac{\partial^2\delta K_2}{\partial
t^2}+K_{22}\delta K_2=0 \qquad .
\end{eqnarray*}
 Therefore, the eigenfunctions of the differential operator
\begin{equation}
K =\left(\begin{array}{cc} K_{11} & 0 \\ 0 & K_{22}
\end{array}\right) =\left(
\begin{array}{cc}
-\frac{d^2}{dx^2}+1-\frac{2\sigma^2}{\cosh^2 \sigma x} & 0 \\
0 & -\frac{d^2}{dx^2}+\sigma^2-\frac{2\sigma^2}{\cosh^2\sigma x}
\end{array} \right) \label{2odo}
\end{equation}
provide the general solution of the linearized equations via the
separation ansatz: $\delta K_1(t,x)=g_1(t)\xi^1(x)$, $\delta
K_2(t,x)=g_2(t)\xi^2(x)$. The eigenvalues and eigenfunctions of $K$
are shown in the following Table:
\begin{center}
\begin{tabular}{|l|c||l|c|} \hline
Eigenvalues &  Eigenfunctions & Eigenvalues & Eigenfunctions \\[0.4cm]
\hline \rule{0.cm}{0.8cm} $\varepsilon_{1-\sigma^2}^2=1-\sigma^2$ &
 $\left(\begin{array}{c} f_{1-\sigma^2}(x)\\ 0 \end{array}\right)
 =\left(\begin{array}{c}\frac{1}{{\rm cosh}^2 \sigma x}\\ 0
\end{array}\right)$ & $\varepsilon_0^2=0$ &
 $\left(\begin{array}{c} 0 \\
f_0(x)\end{array}\right)=\left(\begin{array}{c} 0 \\ {1\over {\rm
cosh}^2\sigma x} \end{array}\right)$\\[0.4cm] $\varepsilon_1^2(q)=\sigma^2 q^2+1$
 & $\left(\begin{array}{c} f_q^1(x)=e^{iq \sigma x}({\rm tanh}\sigma
x-ik)\\ 0 \end{array}\right)$ & $\varepsilon_2^2(k)=\sigma^2(k^2+1)$
 &  $\left(\begin{array}{c} 0 \\
f_k^2(x)=e^{ik\sigma x}({\rm tanh}\sigma x-ik)\end{array}\right)$ \\[0.4cm] \hline
\end{tabular}
\end{center}

\vspace{0.4cm} The spectrum of $K_{22}$ contains a bound state of
zero eigenvalue -the translational mode- and a branch of the
continuous spectrum, with the threshold at
$\varepsilon_2^2(0)=\sigma^2$. ${\rm Spec} K_{11}$ also is formed by
a bound state of positive eigenvalue and a branch of the continuous
spectrum starting at $\varepsilon_1^2(0)=1$. Periodic boundary
conditions in the $[-\frac{l}{2},\frac{l}{2}]$ interval require:
\[
\sigma q\cdot l+\delta_1(q)=2\pi n_1 \qquad , \qquad \sigma k\cdot
l+\delta_2(k)=2\pi n_2 \qquad, \qquad n_1,n_2\in{\mathbb Z} \qquad ,
\]
such that the phase shifts and the induced spectral densities are:
\[
\delta_1(q)=2{\rm arctan}\frac{1}{q}=\delta(q) \qquad , \qquad
\delta_2(k)=2{\rm arctan}\frac{1}{k}=\delta(k)
\]
\begin{equation}
\rho_{K_{11}}(q)=\frac{1}{2\pi}\left(\sigma
l+\frac{d\delta_1}{dq}(q)\right) \qquad , \qquad
\rho_{K_{22}}(k)=\frac{1}{2\pi}\left(\sigma
l+\frac{d\delta_2}{dk}(k)\right) \qquad . \label{ksd}
\end{equation}
In sum, $K$ also acts in the Hilbert space
$L^2=\bigoplus_{\alpha=1}^2 L^2_\alpha(S^1)$, and its spectral
density in the limit of very large radius of the circle is:
\[
\rho_{K}(k)=\left(\begin{array}{cc} \frac{dn_1}{dk} & 0
\\ 0 & \frac{dn_2}{dq} \end{array}\right)={1\over 2\pi}\left(
\sigma l+\frac{d\delta}{dk}(k)\right) \left(\begin{array}{cc} 1 & 0
\\ 0 & 1 \end{array}\right) \qquad .
\]
\section{One-loop shift to the classical $K$ kink
masses in the massive non-linear $S^2$-sigma model}

\subsection{Zero-point kink energy}

The general solution of the linearized field equations governing the
small kink fluctuations is:
\begin{eqnarray*}
\delta
K_1(x_0,x)&=&\frac{1}{2}\cdot\frac{1}{\sqrt{2\sqrt{1-\sigma^2}}}\left(A_{1-\sigma^2}
e^{i\sqrt{1-\sigma^2}x_0}+A_{1-\sigma^2}^*e^{-i\sqrt{1-\sigma^2}x_0}\right)
f_{1-\sigma^2}(x)\\&+&\frac{1}{2}\cdot\sqrt{{1\over
l}}\sum_k{1\over\sqrt{2\varepsilon_1(k)}}
\left\{A_1(k)e^{-i\varepsilon_1(k)x_0}f_k^1(x)+A_1^*(k)e^{i\varepsilon_1(k)x_0}f_k^{*1}(x)\right\}
\\ \delta
K_2(x_0,x)&=&\frac{1}{2}\cdot\sqrt{{1\over
l}}\sum_k{1\over\sqrt{2\varepsilon_2(k)}}
\left\{A_2(k)e^{-i\varepsilon_2(k)x_0}f_k^2(x)+A_2^*(k)e^{i\varepsilon_2(k)x_0}f_k^{*2}(x)\right\}
\qquad .
\end{eqnarray*}
Note that the zero mode is not included because only contribute to
quantum corrections at two-loop order. In the orthogonal complement
to the kernel of $K_{22}$ in
$\bigoplus_{\alpha=1}^2L^2_\alpha({\mathbb S}^1)$, the
eigenfunctions of the $K$ operator satisfying PBC form a complete
orthonormal system. Therefore, the classical free Hamiltonian
\begin{eqnarray*}
H^{(2)}&=&E+\int_{-\frac{l}{2}}^{\frac{l}{2}} \, dx \,
\left\{{\lambda\over 2}\sum_{\alpha=1}^2\left(\frac{\partial\delta
K_\alpha}{\partial x_0}\cdot\frac{\partial\delta K_\alpha}{\partial
x_0}+\delta
K_\alpha\cdot K_{\alpha\alpha}\cdot \delta K_\alpha\right)\right\}\\
&=& E+{\lambda\over 2}\left[
\sqrt{1-\sigma^2}\left(A_{1-\sigma^2}^*A_{1-\sigma^2}
+A_{1-\sigma^2} A_{1-\sigma^2}^*\right)
+\sum_k\sum_{\alpha=1}^2\left\{\varepsilon_\alpha(k)(A_\alpha^*(k)A_\alpha(k)+
A_\alpha(k)A_\alpha^*(k))\right\}\right]
\end{eqnarray*}
can be written in terms of the normal modes of the system in the
quadratic approximation. From this expression, one passes via
canonical quantization,
$[\hat{A}_\alpha(k),\hat{A}^\dagger_\beta(q)]=\delta_{\alpha\beta}\delta_{kq}$,
$[\hat{A}_{1-\sigma^2},\hat{A}_{1-\sigma^2}^\dagger]=1$, to the
quantum free Hamiltonian:
\[
\hat{H}^{(2)}=E+\lambda\sqrt{1-\sigma^2}\left(\hat{A}_{1-\sigma^2}^\dagger\hat{A}_{1-\sigma^2}+{1\over
2}\right)+\lambda\sum_k \, \sum_{\alpha=1}^2 \, \left[
\varepsilon_\alpha(k)\left(\hat{A}_\alpha^\dagger(k)\hat{A}_\alpha(k)+{1\over
2}\right)\right] \qquad .
\]
The kink ground state is a coherent state annihilated by all the
destruction operators:
\[
\hat{A}_\alpha(k)|0;K\rangle =\hat{A}_{1-\sigma^2}|0;K \rangle =0 \,
\, , \forall k \, , \, \forall \alpha \nonumber \qquad , \qquad
\hat{\phi}_a(t,x)|0;K \rangle = \phi_a^K(x)|0;K \rangle \qquad .
\]
The kink ground state energy is:
\begin{equation}
E+\Delta E=\langle 0;K |\hat{H}^{(2)}|0;K \rangle =2\lambda R^2
\sigma+{\lambda\over 2}\sqrt{1-\sigma^2}+{\lambda\over 2}\sum_k\,
\sum_{\alpha=1}^2\,\varepsilon_\alpha(k)=2\lambda R^2
\sigma+{\lambda\over 2}{\rm Tr}_L^2 K^{{1\over 2}} \qquad .
\label{olms}
\end{equation}

\subsection{Zeta function regularization and Casimir kink energy}

Both ${\rm Tr}_{L^2}K_0^{1\over 2}$ and ${\rm Tr}_{L^2}K^{1\over 2}$
are ultraviolet divergent quantities: one sums over an infinite
number of eigenvalues, and a regularization/renormalization
procedure must be implemented to make sense of these formal
expressions. We renormalize the zero-point kink energy by
subtracting from it the vacuum energy to define the kink Casimir
energy:
\[
\bigtriangleup E^C=\bigtriangleup E-\bigtriangleup
E_0=\frac{\lambda}{2}\left[{\rm Tr}_{L^2}K^{1\over 2}-{\rm
Tr}_{L^2}K_0^{1\over 2}\right] \qquad .
\]
The subtraction of these two divergent quantities is regularized by
using the associated generalized zeta functions, i.e., we
temporarily assign to $\bigtriangleup E^C$ the finite value:
\[
\bigtriangleup
E^C(s)=\frac{\mu}{2}\left(\frac{\mu^2}{\lambda^2}\right)^s\left[{\rm
Tr}_{L^2}K^{-s}-{\rm
Tr}_{L^2}K_0^{-s}\right]=\frac{\mu}{2}\left(\frac{\mu^2}{\lambda^2}\right)^s\left[\zeta_K(s)-\zeta_K(s)\right]
\]
at a regular point of both $\zeta_K(s)$ and $\zeta_{K_0}(s)$. Here,
\[
\zeta_K(s)=\sum_{{\rm Spec}K}\, \lambda^{-s} \quad , \quad
\zeta_{K_0}(s)=\sum_{{\rm Spec}K_0}\, \lambda_0^{-s} \qquad ; \qquad
\lambda \in {\rm Spec}K \, \, , \, \, \lambda_0 \in {\rm Spec}K_0
\quad , \quad s\in{\mathbb C}
\]
are the spectral zeta functions of $K$ and $K_0$, which are
meromorphic functions of the complex variable $s$. An auxiliary
parameter $\mu$ with dimensions of inverse length is used to keep
the physical dimension right and we shall go to the physical limit
$\Delta E^C=\lim_{s\rightarrow -{1\over 2}} \Delta E^C(s)$ at the
end of the process.

\subsection{Partition and generalized zeta functions}

Because analytical information about the spectrum of $K$ is only
available at the the limit of large $l$ (bound state energies, phase
shifts and spectral densities) it is better to consider first the
partition or heat functions:
\[
{\rm Tr}_{L^2}\, e^{-\beta K_0} = \frac{\sigma l}{2\pi}
\left(\int_{-\infty}^\infty dk e^{-(\sigma^2 k^2+1)\beta}
+\int_{-\infty}^\infty dk e^{-\sigma^2 (k^2+1)\beta} \right) =
\frac{l}{\sqrt{4\pi \beta}} (e^{-\beta}+e^{-\sigma^2\beta}) \quad ,
\quad \beta\in{\mathbb R} \qquad .
\]
Note that here we have replaced $k$ and $q$ defined in Section \S
3.1 by $\sigma k$ and $\sigma q$ for a better comparison between the
spectra of $K_0$ and $K$. The PBC spectral density of $K_0$ is thus
obtained by replacing $\lambda$ by $\gamma$. The $K$-heat function
is also expressed in terms of integrals over the continuous spectrum
at the $l=\infty$ limit, rather than infinite sums. The integrals,
however, must be weighted with the PBC spectral densities:
\begin{eqnarray*}
{\rm Tr}^*_{L^2}\,e^{-\beta K}&=& {\rm Tr}_{L^2}\,e^{-\beta K_0}
+e^{-(1-\sigma^2)\beta}+\frac{1}{2\pi}\int_{-\infty}^\infty \, dk \,
\frac{d\delta}{dk}\left[e^{-(\sigma^2 k^2+1)\beta}+e^{-\sigma^2
(k^2+1)\beta}\right]\\&=& {\rm Tr}_{L^2}\,e^{-\beta K_0} +
e^{-(1-\sigma^2)\beta} {\rm Erf}\,(\sigma \sqrt{\beta})- {\rm
Erfc}(\sigma \sqrt{\beta}) \qquad .
\end{eqnarray*}
The error and complementary error functions of
$\beta=\frac{\lambda}{k_B T}$, a fictitious inverse temperature,
arise and the asterisk means that we have not included the zero mode
because zero modes do not enter the one-loop formula (\ref{olms}).

The generalized zeta functions are Mellin transforms of the heat
functions:
\begin{eqnarray*}
\zeta_{K_0}(s)&=& \frac{1}{\Gamma(s)} \int_0^\infty d\beta \,
\beta^{s-1} \, {\rm Tr}_{L^2}\, e^{-\beta K_0} = \frac{\sigma
l}{2\pi}\int_{-\infty}^\infty \, dk \, \left(\frac{1}{[\sigma^2
k^2+1]^s}+\frac{1}{[\sigma^2 k^2+\sigma^2]^s}\right)=\frac{l}{
\sqrt{4\pi}} \frac{\Gamma(s-\frac{1}{2})}{\Gamma(s)} \left(1+
\frac{1}{\sigma^{2s-1}}\right)\\
\zeta_K^*(s)&=&\frac{1}{\Gamma(s)} \int_0^\infty d\beta \,
\beta^{s-1} \, {\rm Tr}_{L^2}\, e^{-\beta K} =\zeta_{K_0}(s)+
\frac{\Gamma(s+\frac{1}{2})}{\sqrt{\pi}\Gamma(s)} \left[
\frac{2\sigma}{(1-\sigma^2)^{s+\frac{1}{2}}}
{}_2F_1[{\textstyle\frac{1}{2}},s+
{\textstyle\frac{1}{2}},{\textstyle\frac{3}{2}};
-{\textstyle\frac{\sigma^2}{1-\sigma^2}}] -\frac{1}{s\sigma^{2s}}
 \right] \, \, \, .
\end{eqnarray*}
We indeed find meromorphic functions of $s$ with poles and residues
determined from the poles and residues of Euler $\Gamma(s)$ and
Gauss hypergeometric ${}_2F_1[a,b,c,z]$ functions
{\footnote{Strictly speaking, Mellin transforms are defined in their
fundamental strips, respectively ${\rm Re}s>1/2$, ${\rm Re}s>0$ in
our problems. In the spirit of zeta function regularization, we
extend the results of the Mellin transforms to the whole complex
$s$-plane by analytic continuation.}}.

In the APPENDIX I we show that the kink Casimir energy in the
physical limit $s=-\frac{1}{2}$ is the divergent quantity:
\begin{equation}
\Delta E^C = -\frac{\lambda \sigma}{2\pi} \left[\lim_{\varepsilon
\rightarrow 0} \frac{2}{\varepsilon} + 2 \ln \frac{\mu^2}{\lambda^2}
+ \ln \frac{16}{\sigma^2(1-\sigma^2)} -2 +
{}_2F_1^{(0,1,0,0)}[\frac{1}{2},0,\frac{3}{2},-\frac{\sigma^2}{1-\sigma^2}]
\right] \label{kcen} \qquad ,
\end{equation}
where
${}_2F_1^{(0,1,0,0)}[\frac{1}{2},0,\frac{3}{2},-\frac{\sigma^2}{1-\sigma^2}]$
is the derivative of the Gauss hypergeometric function with respect
to the second argument.
\subsection{Zeta function regularization of the self-energy graphs and
kink mass renormalization} It remains to take the effect of mass
renormalization into account. The contribution to the kink energy
due to the mass renormalization counter-terms is:
\[
\bigtriangleup E^{MR}=-\lambda \int \, dx \, {\cal
L}_{C.T.}(\phi_1^{{\rm K}_1},\phi_2^{{\rm
K}_1})=\lambda\frac{\sigma^2}{R^2}[I(1)+I(\sigma^2)]\int \, dx \,
\phi_2^{{\rm K}_1}(x)\phi_2^{{\rm K}_1}(x)=2\lambda\sigma
[I(1)+I(\sigma^2)] \qquad .
\]
In the normalization interval of length $l$ the integrals become
infinite sums
\[
I(1)=\frac{\sigma}{2}\int \, \frac{dk}{2\pi}\,
\frac{1}{\sqrt{\sigma^2 k^2+1}}=\frac{1}{2l}\sum_{n=-\infty}^\infty
\, \frac{1}{(\sigma^2 n^2+1)^{1\over 2}}\quad , \quad
I(\sigma^2)=\frac{\sigma}{2}\int \, \frac{dk}{2\pi}\,
\frac{1}{\sqrt{\sigma^2
k^2+\sigma^2}}=\frac{1}{2l}\sum_{n=-\infty}^\infty \,
\frac{1}{(\sigma^2 n^2+\sigma^2)^{1\over 2}}
\]
that can be regularized by using zeta functions:
\[
I(1)= -\frac{1}{\mu L} \lim_{s\rightarrow -\frac{1}{2}}
(\frac{\mu^2}{\lambda^2})^{s+1} \frac{\Gamma(s+1)}{\Gamma(s)}
\zeta_{K_{011}}(s+1) \quad , \quad  I(\sigma^2)= -\frac{1}{\mu L}
\lim_{s\rightarrow -\frac{1}{2}} (\frac{\mu^2}{\lambda^2})^{s+1}
\frac{\Gamma(s+1)}{\Gamma(s)} \zeta_{K_{022}}(s+1) \, \, \, ,
\]
such that {\footnote{The differential operators $K_{011}$ and
$K_{022}$ are defined in Page 4.}}:
\[
\Delta E^{MR}(s)=-\frac{2\sigma \lambda^2}{\mu\sqrt{4\pi}}
(\frac{\mu^2}{\lambda^2})^{s+1}
\frac{\Gamma(s+\frac{1}{2})}{\Gamma(s)}  \left(1+
\frac{1}{\sigma^{2s+1}} \right) \qquad .
\]
In the APPENDIX I it is proved that the physical limit $s=-{1\over
2}$ is also a pole of $\Delta E^{MR}(s)$:
\begin{equation}
\Delta E^{MR}= \frac{\lambda \sigma}{2\pi} \left[
\lim_{\varepsilon\rightarrow 0} \frac{2}{\varepsilon}+ 2 \ln
\frac{\mu^2}{\lambda^2} +2(\ln 4-2)- \ln \sigma^2 \right]
\label{kmre} \qquad  .
\end{equation}
The divergent terms in $\Delta E^C$ (\ref{kcen}) and $\Delta E^{MR}$
(\ref{kmre}), as well as the $\mu$-dependent terms, cancel each
other exactly and the one-loop $K$ kink mass shift is:
\begin{equation}
\Delta
E=-\frac{\lambda\sigma}{2\pi}\left[2+{}_2F_1^{(0,1,0,0)}[\frac{1}{2},0,\frac{3}{2},-\frac{\sigma^2}{1-\sigma^2}]
-\ln(1-\sigma^2)\right]=-\frac{\lambda\sigma}{\pi}[2-\frac{\sqrt{1-\sigma^2}}{\sigma}{\rm
arccos}\sqrt{1-\sigma^2}\, \, \, ] \label{olmk1} \qquad .
\end{equation}
In formula (\ref{olmk1}) we have also written the result found in
our derivation à la Cahill-Comtet-Glauber of the quantum correction,
see \cite{AMAJ1}. The heat kernel/zeta function result is
$-\frac{\lambda\sigma}{\pi}f(\sigma)$ whereas the CCH formula leads
to $-\frac{\lambda\sigma}{\pi}g(\sigma)$, where
\[
f(\sigma)=1+\frac{1}{2}{}_2F_1^{(0,1,0,0)}[\frac{1}{2},0,\frac{3}{2},-\frac{\sigma^2}{1-\sigma^2}]
-\frac{1}{2}\ln(1-\sigma^2) \qquad , \qquad
g(\sigma)=2-\frac{\sqrt{1-\sigma^2}}{\sigma}{\rm
arccos}\sqrt{1-\sigma^2} \qquad .
\]
Despite appearances, $f(\sigma)$ and $g(\sigma)$ are identical
functions of $\sigma\in [0,1]$, as the Mathematica plots in the
Figure 1 show. \noindent\begin{figure}[htbp]
\centerline{\epsfig{file=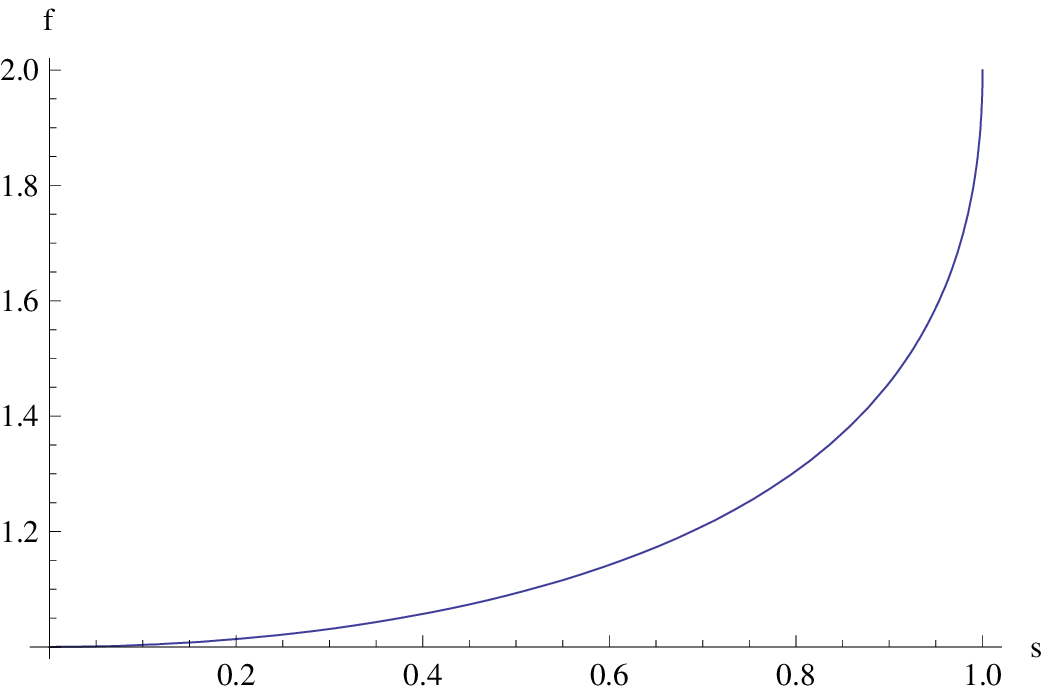, height=3.5cm}\quad
\epsfig{file=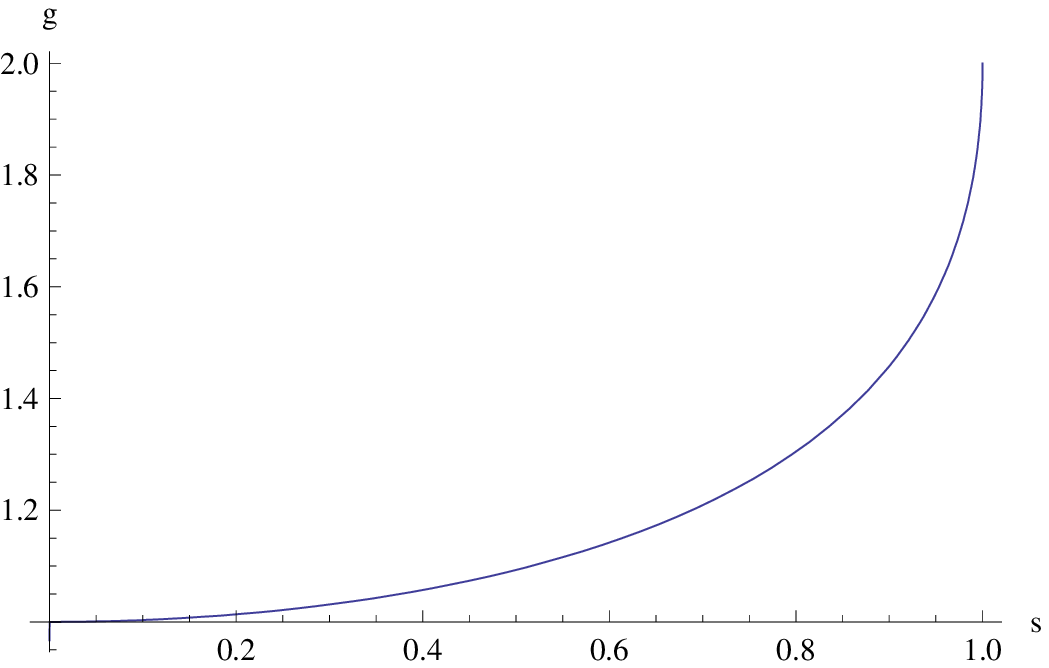, height=3.5cm} \quad
\epsfig{file=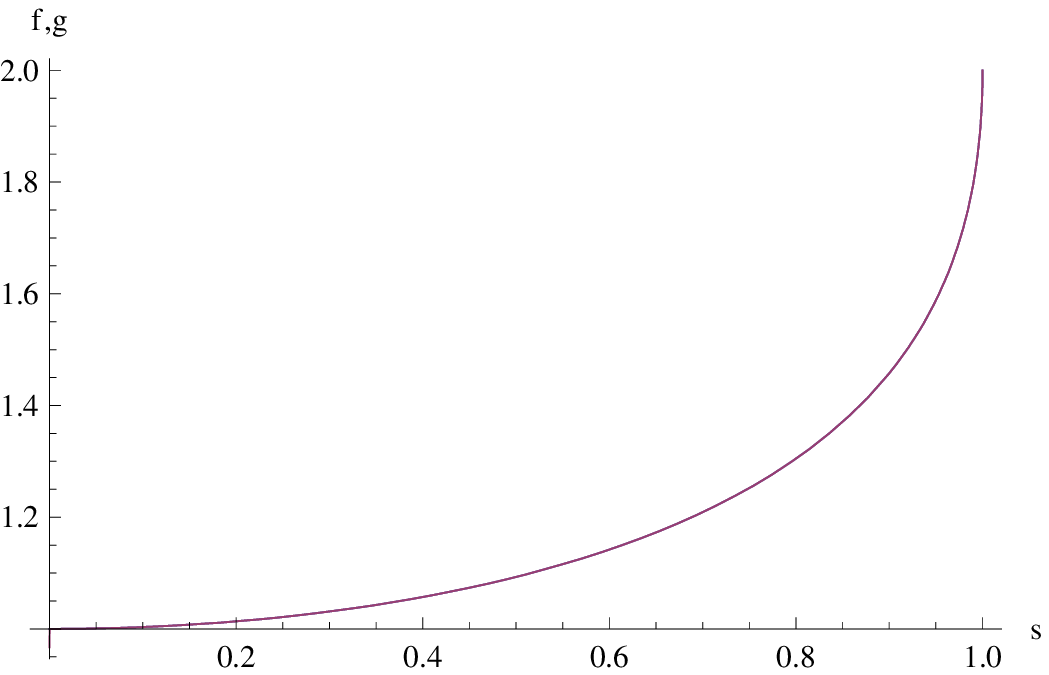, height=3.5cm} }\caption{\small {\it
Graphics of $f(\sigma)$ (left), $g(\sigma)$ (center), and
$f(\sigma)$ and $g(\sigma)$ plotted together (right). In the Figure,
$\sigma$ is labeled as $s$ in the abscissa axis. }}
\end{figure}
This is remarkable: there is no mention about the analytic identity
between the functions $f(\sigma)$ and $g(\sigma)$ in the ample
Literature on special functions. Nevertheless, they trace identical
curves as functions of $\sigma$.

\section{High-temperature asymptotic expansion}

The exact heat or partition function can be written in the form:
\[
{\rm Tr}_{L^2}e^{-\beta K}={\rm tr}\left(\begin{array}{cc} {\rm
Tr}_{L_1^2}e^{-\beta K_{11}} & 0 \\ 0 & {\rm Tr}_{L_2^2}e^{-\beta
K_{22}}\end{array}\right)=\left(\frac{l}{\sqrt{4\pi\beta}}+e^{\sigma^2\beta}{\rm
Erf}[\sigma\sqrt{\beta}]\right){\rm tr}\left(\begin{array}{cc}
e^{-\beta} & 0 \\ 0 & e^{-\sigma^2\beta}\end{array}\right) \qquad ,
\]
where $\lq\lq {\rm tr}"$ means trace in the matrix sense. There is
an alternative way of computing this quantity by means of a
high-temperature asymptotic expansion. Although we have the exact
formula in our system, we shall also perform the approximate
calculation, which is the only one possible in other systems in
order to gain control of this second approach in this favorable
case.

 In the APPENDIX II it is shown how the coefficients
 of the power expansion of the $K$-heat trace
\begin{equation}
{\rm Tr}e^{-\beta K}=\frac{1}{\sqrt{4\pi}}\sum_{n=0}^\infty \,
c_n(K)\beta^{n-\frac{1}{2}}{\rm tr}\left(\begin{array}{cc}
e^{-\beta} & 0 \\ 0 & e^{-\sigma^2\beta}\end{array}\right)\qquad ,
\label{eq:serie}
\end{equation}
the Seeley coefficients $c_n(K)$, are obtained through integration
of the Seeley densities over the whole line. The densities satisfy
recurrence relations tantamount to the heat kernel equation starting
from a general potential $U(x)$. In our problem we must solve the
recurrence relations between these densities for the potential
$U(x)=-\frac{2\sigma^2}{\cosh^2 \sigma x}$, essentially the same
potential as for the sine-Gordon kink, see \cite{AMAJW0}. We list
these coefficients up to the twentieth order in Table IV:

\begin{table}[htb]
\begin{tabular}{|c|c|}
\hline $n$ & $c_n(K)/\sigma^{2n-1}$ \\ \hline
1  & $4.$\\
2  & $2.66667$ \\
3  & $1.06667$ \\
4  & $0.304762$ \\
5  & $0.0677249$ \\
6  & $0.0123136$ \\
7  & $1.8944 \times 10^{-3}$  \hspace{0.5cm}\\ \hline
\end{tabular}\hspace{1.5cm} \begin{tabular}{|c|c|}
\hline $n$ & $c_n(K)/\sigma^{2n-1}$ \\ \hline
8  & $2.52587 \times 10^{-4}$ \\
9  & $2.97161 \times 10^{-5}$ \\
10 & \hspace{0.5cm} $3.12801 \times 10^{-6}$\\
11  & $2.97906 \times 10^{-7}$\\
12  & $2.59049 \times 10^{-8}$ \\
13  & $2.072239 \times 10^{-9}$ \\
14  & $1.5351 \times 10^{-10}$ \hspace{0.5cm}\\
\hline
\end{tabular}
\hspace{1.5cm} \begin{tabular}{|c|c|} \hline $n$ &
$c_n(K)/\sigma^{2n-1}$ \\ \hline
15  & $1.05869 \times 10^{-11}$ \\
16  & $6.83027 \times 10^{-13}$ \\
17  & $4.13956 \times 10^{-14}$ \\
18  & $2.36546 \times 10^{-15}$ \\
19  & $1.27863 \times 10^{-16}$ \\
20  & \hspace{0.5cm}$6.55706 \times 10^{-18}$\hspace{0.5cm} \\
\hline
\end{tabular}
\caption{Seeley Coefficients}
\end{table}
Write now the spectral zeta functions in the form:
\begin{eqnarray*}
\zeta_{K_0}(s)&=&\zeta_{K_0}(s;b)+B_{K_0}(s;b)\\
&=&\frac{1}{\Gamma(s)}\cdot\frac{l}{\sqrt{4\pi}}\, \left[{\rm tr}\,
\int_0^b\, d\beta \, \beta^{s-\frac{3}{2}}\,\left(\begin{array}{cc}
e^{-\beta} & 0 \\ 0 & e^{-\sigma^2\beta}
\end{array}\right)+
{\rm tr}\, \int_b^\infty\, d\beta \, \beta^{s-\frac{3}{2}}\,
\left(\begin{array}{cc} e^{-\beta} & 0 \\ 0 & e^{-\sigma^2\beta}
\end{array}\right)\right]\\&=&
\frac{1}{\Gamma(s)}\cdot\frac{l}{\sqrt{4\pi}}\,\left[{\rm
tr}\left(\begin{array}{cc}\gamma[s-\frac{1}{2},b] & 0
\\ 0 & \frac{\sigma}{\sigma^{2s}}\gamma[s-\frac{1}{2},\sigma^2
b]\end{array}\right)+{\rm
tr}\left(\begin{array}{cc}\Gamma[s-\frac{1}{2},b] & 0
\\ 0 & \frac{\sigma}{\sigma^{2s}}\Gamma[s-\frac{1}{2},\sigma^2
b]\end{array}\right)\right]
\end{eqnarray*}
\begin{eqnarray*}
\zeta_{K}(s)&=&\zeta_{K}(s;b)+B_{K}(s;b)\\
&=&\frac{1}{\Gamma(s)\sqrt{4\pi}}\,\sum_{n=0}^\infty \, c_n(K)\, \,
{\rm tr}\, \int_0^b\, d\beta \,
\beta^{s+n-\frac{3}{2}}\,\left(\begin{array}{cc} e^{-\beta} & 0 \\ 0
& e^{-\sigma^2\beta}
\end{array}\right)+ \frac{1}{\Gamma(s)}
\int_b^\infty \, d\beta \, \beta^{s-1} \, {\rm Tr}_{L^2}\, e^{-\beta
K}\\&=& \frac{1}{\Gamma(s)\sqrt{4\pi}}\, \sum_{n=0}^\infty \, c_n(K)
\, \, {\rm tr}\left(\begin{array}{cc}\gamma[s+n-\frac{1}{2},b] & 0
\\ 0 & \frac{\sigma}{\sigma^{2(s+n)}}\gamma[s+n-\frac{1}{2},\sigma^2
b]\end{array}\right)+\frac{1}{\Gamma(s)}\int_b^\infty \, d\beta \,
\beta^{s-1}\, {\rm Tr}_{L^2}\, e^{-\beta K}\, \, .
\end{eqnarray*}
The incomplete Euler Gamma functions $\gamma[z,a]$ are meromorphic
functions of $z$ whereas $B_{K_0}(s;b)$ and $B_K(s;b)$ are entire
functions of $s$. The splitting point of the Mellin transform is
usually taken at $b=1$. We leave $b$ as a free parameter for reasons
to be explained later.

Neglecting the entire parts, the zero-point energy renormalization
\[
\zeta_K(s;b)-\zeta_{K_0}(s;b)=\frac{1}{\Gamma(s)\sqrt{4\pi}}\sum_{n=1}^\infty\,
c_n(K)\,{\rm tr}\left(\begin{array}{cc} \gamma[s+n-\frac{1}{2},b] &
0
\\ 0 &
\frac{\sigma}{\sigma^{2(s+n)}}\gamma[s+n-\frac{1}{2},\sigma^2
b]\end{array}\right)
\]
gets rid of the $c_0(K)$ term. The contribution of $c_1(K)$
\[
\bigtriangleup E^C_{(1)}=\frac{1}{\sqrt{\pi}} \lim_{s\rightarrow
-\frac{1}{2}}\left(\frac{\mu^2}{\lambda^2}\right)^s\frac{\mu}{\Gamma(s)}\,
{\rm tr}\left(\begin{array}{cc} \sigma\gamma[s+\frac{1}{2},b] & 0
\\ 0 & \frac{1}{\sigma^{2s}}\gamma[s+\frac{1}{2}, \sigma^2 b]\end{array}\right)
\]
is exactly canceled by the mass renormalization counter-terms:
\[
\bigtriangleup E^{\rm MR}=-\frac{1}{\pi}\lim_{s\rightarrow
-\frac{1}{2}}\left(\frac{\mu^2}{\lambda^2}\right)^{s+1}\frac{\sigma\lambda^2}{\mu\Gamma(s)}\,
{\rm tr}\, \left(\begin{array}{cc} \gamma[s+\frac{1}{2},b] & 0
\\ 0 & \frac{1}{\sigma^{2s+1}}\gamma[s+\frac{1}{2},\sigma^2
b]\end{array}\right) \qquad .
\]
We must now subtract the contribution of the zero mode:
\begin{eqnarray*}
\zeta^*_K(s;b)&=&\zeta_K(s;b)-\frac{1}{\Gamma(s)}\lim_{\varepsilon\to
0}\, \int_0^b \, d\beta \, \beta^{s-1}\, e^{-\varepsilon \beta}\\
&=& \zeta_K(s;b)-\frac{1}{\Gamma(s)}\lim_{\varepsilon\to
0}\,\frac{1}{\varepsilon^s}\gamma[s,\varepsilon
b]=\zeta_K(s;b)-\frac{b^s}{s\Gamma(s)}
\end{eqnarray*}
Finally, the high-temperature one-loop correction to the $K$ kink
energy is:
\[
\bigtriangleup E(b)=\frac{\mu}{2}\lim_{s\rightarrow
-\frac{1}{2}}\left(\frac{\mu^2}{\lambda^2}\right)^{s}\frac{1}{\Gamma(s)}
\left(\frac{1}{\sqrt{4\pi}}\sum_{n=2}^\infty\, c_n(K)\, {\rm
tr}\left(\begin{array}{cc} \gamma[s+n-\frac{1}{2},b] & 0 \\ 0 &
\frac{\sigma}{\sigma^{2(s+n)}}\gamma[s+n-\frac{1}{2},\sigma^2
b]\end{array}\right) -\frac{b^s}{s}\right) \qquad .
\]
In practice, truncation of the series is also necessary:
\begin{equation}
\bigtriangleup
E(b,N_0)=-\frac{\lambda}{4\sqrt{\pi}}\left[\frac{2}{\sqrt{b}}+\frac{1}{\sqrt{4\pi}}\sum_{n=2}^{N_0}\,
c_n(K)\, {\rm tr} \, \left(\begin{array}{cc} \gamma[n-1,b] & 0
\\ 0 & \frac{\sigma^2}{\sigma^{2n}}\gamma[n-1,\sigma^2
b]\end{array}\right)\right] \label{olmsf} \qquad .
\end{equation}
Using formula (\ref{olmsf}) to calculate the one-loop kink mass
shift, we admit an error of:
\[
\bigtriangleup E-\bigtriangleup
E(b,N_0)=-\frac{\lambda}{2^3\pi}\left(\sum_{n=2}^{N_0}\,
c_n(K)\left(\Gamma[n-1,b]+\frac{\Gamma[n-1,\sigma^2
b]}{\sigma^{2(n-1)}}\right)+\sum_{n=N_0+1}^{\infty}\,
c_n(K)\Gamma(n-1)\left(1+\frac{1}{\sigma^{2(n-1)}}\right)\right)
\qquad .
\]
We offer a Figure where formula (\ref{olmsf}) has been applied for
$N_0=20$ and several values of $\sigma$. The very good precision of
the asymptotic formula was achieved by adapting the parameter $b$ to
the value of $\sigma$. For instance, we have taken $b=1000$ for
$\sigma=0.1$, $b=100$ for $\sigma=0.3$, $b=50$ for $\sigma=0.5$,
$b=20$ for $\sigma=0.7$, $b=10$ for $\sigma=0.9$, and $b=10$ for
$\sigma=1$. Physically, this means that the lighter the particle
mass ($\sigma^2$) is, the longer the integration interval in the
Mellin transform must be taken to minimize the error produced by the
neglected entire parts. In practice, we have chosen $b$ in each case
at the frontier near the point $\beta_0\in(0,\infty]$, where the
asymptotic formula of the $K$-heat trace departs from its exact
value.

\noindent\begin{figure}[htbp] \centerline{\epsfig{file=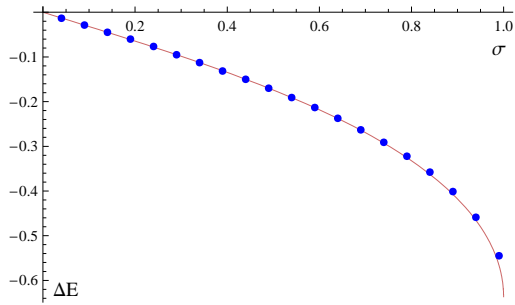,
height=4cm} }\caption{\small {\it Points obtained using the
asymptotic expansions for several values of $\sigma $ plotted on the
exact curve giving the one-loop mass shift as a function of $\sigma
$. }}
\end{figure}

\section{Conclusions and further comments}

In sum, we may draw the following conclusions:
\begin{enumerate}
\item We have obtained the one-loop mass shift to the classical mass of
the stable topological kink that exists in a massive anisotropic
non-linear ${\mathbb S}^2$-sigma model.

\item In the isotropic case, $\sigma=1$, our result agrees with the
answer provided by other authors: the one-loop correction is twice
(in modulus) the correction for the sine-Gordon kink, see
\cite{MRvNW} and \cite{RSvN}.

\item Our procedure is based on the heat kernel/zeta function
regularization method. The result is identical to the answer
achieved by means of the Cahill-Comtet-Glauber formula.

This is a remarkable fact: the CCH formula takes into account only
the bound state eigenvalues and the thresholds to the two branches
of the continuous spectrum of the Schr$\ddot{\rm o}$dinger operators
that govern the field small fluctuations. It is essentially finite.
Our computation involves infinite renormalizations. The criterion
chosen to set finite renormalizations -no modification of the
particle masses at the one-loop level, equivalent to the vanishing
tadpole criterion in linear sigma models- does exactly the job.

\item We have also derived a high-temperature approximated formula
for the mass shift, relying on the heat kernel asymptotic expansion.
We stress that we have improved a former weakness of our method. The
approximation to the exact result was poor for light masses
-non-dimensional mass $<1$- in the model studied in \cite{AMAJ1}. We
have achieved a very good approximation in this paper even for light
particles by enlarging the integration interval of the Mellin
transform and considering an optimum number of Seeley coefficients.
We believe that this is a general procedure, working also in models
where the exact generalized zeta function is not available.
\end{enumerate}

As a final comment, we look forward to addressing the quantization
procedure for: (a) Multi-solitons and breather modes of this model.
(b) Stable topological kinks that may arise in other massive
non-linear sigma models with different potentials, e.g., quartic,
and/or different target manifolds, e.g., ${\mathbb S}^3$.

\section{Acknowledgements}

We thank the Spanish Ministerio de Educacion y Ciencia and Junta de
Castilla y Leon for partial financial support under grants
FIS2006-09417, GR224, and SA034A08. JMG thanks the ESF Research
Network CASIMIR for providing excellent opportunities on the Casimir
effect and related topics like topological defect fluctuations.

\section*{APPENDIX I: Kink Casimir energy and mass renormalization near the pole}

The Casimir kink energy is, see Section \S. V.C:
\begin{eqnarray*}
\Delta E^C = \lim_{s\rightarrow -\frac{1}{2}} \Delta E^C(s) &=&
\lim_{s\rightarrow -\frac{1}{2}} \left[ \frac{\mu}{2\sqrt{\pi}}
\left(\frac{\mu^2}{\lambda^2}\right)^s
\frac{\Gamma(s+\frac{1}{2})}{\Gamma(s)} \left(
\frac{2\sigma}{(1-\sigma^2)^{s+\frac{1}{2}}} \,
{}_2F_1[{\textstyle\frac{1}{2}},s+{\textstyle\frac{1}{2}},{\textstyle\frac{3}{2}};
-{\textstyle\frac{\sigma^2}{1-\sigma^2}}]-\frac{1}{s\sigma^{2s}}
 \right) \right] \\
 &=& \frac{\lambda \sigma}{2\sqrt{\pi}} \lim_{\varepsilon \rightarrow 0}
 \left[  \left(\frac{\mu^2}{\lambda^2}\right)^\varepsilon
 \frac{\Gamma(\varepsilon)}{\Gamma(-\frac{1}{2}+\varepsilon)} \left(
\frac{2}{(1-\sigma^2)^{\varepsilon}} \,
{}_2F_1[{\textstyle\frac{1}{2}},\varepsilon,{\textstyle\frac{3}{2}};
{\textstyle\frac{\sigma^2}{1-\sigma^2}}]-
\frac{1}{(-\frac{1}{2}+\varepsilon) \sigma^{2\varepsilon}}
 \right) \right] \qquad ,
\end{eqnarray*}
but $s=-{1\over 2}$ is a pole of $\Delta E^C(s)$. To find the
residue, we expand this function in the neighborhood of the pole by
using the following results
\begin{eqnarray*}
\left(\frac{\mu^2}{\lambda^2}\right)^\varepsilon
\frac{\Gamma(\varepsilon)}{\Gamma(-\frac{1}{2}+\varepsilon)}&
\simeq& -\frac{1}{2\sqrt{\pi}} \left( \frac{1}{\varepsilon} + \ln
\frac{\mu^2}{\lambda^2} +\ln 4 -2 \right) \quad , \quad \qquad
\frac{1}{-\frac{1}{2}+\varepsilon} \frac{1}{\sigma^{2\varepsilon}}
\simeq  -2-\varepsilon (4-2\ln \sigma^2) \\
\frac{2}{(1-\sigma^2)^\varepsilon} \,
{}_2F_1[{\textstyle\frac{1}{2}},\varepsilon,{\textstyle\frac{3}{2}};
{\textstyle\frac{\sigma^2}{1-\sigma^2}}] &\simeq & 2-2 \varepsilon
(\ln (1-\sigma^2)-
{}_2F_1^{(0,1,0,0)}[\frac{1}{2},0,\frac{3}{2},-\frac{\sigma^2}{1-\sigma^2}])
\end{eqnarray*}
where
${}_2F_1^{(0,1,0,0)}[\frac{1}{2},0,\frac{3}{2},-\frac{\sigma^2}{1-\sigma^2}]$
is the derivative of the Gauss hypergeometric function with respect
to the second argument and we made use of the fact that
${}_2F_1[\frac{1}{2},0,\frac{3}{2},-\frac{\sigma^2}{1-\sigma^2}]=1$.

The physical limit $s=-{1\over 2}$ is also a pole of $\Delta
E^{MR}(s)$, see Section \S. V.D:
\begin{eqnarray}
\Delta E^{MR}&=& -\frac{2\sigma \lambda}{\sqrt{4\pi}}
\lim_{\varepsilon\rightarrow 0}
\left(\frac{\mu^2}{\lambda^2}\right)^\varepsilon
\frac{\Gamma(\varepsilon)}{\Gamma(-\frac{1}{2} +\varepsilon)} \left(
\frac{1}{\sigma^{2\varepsilon}}+1 \right)=\frac{\lambda
\sigma}{2\pi} \lim_{\varepsilon \rightarrow 0} \left(1+\varepsilon
\ln\frac{\mu^2}{\lambda^2}\right)\left(\frac{1}{\varepsilon}+\psi(1)\right)
\left(1-\varepsilon \psi(-\frac{1}{2})\right)(2-\varepsilon
\ln\sigma^2)\nonumber\\&=& \frac{\lambda \sigma}{2\pi} \left[
\lim_{\varepsilon\rightarrow 0} \frac{2}{\varepsilon}+ 2 \ln
\frac{\mu^2}{\lambda^2} +2(\ln 4-2)- \ln \sigma^2 \right]
\label{kmre} \qquad  .
\end{eqnarray}

\section*{APPENDIX II: The heat kernel expansion}

Consider the $K_0$- and $K$-heat kernels:
\begin{eqnarray}
&&\left(\frac{\partial}{\partial\beta}+K_0\right)K_{K_0}(x,y;\beta)=0
\qquad , \qquad K_{K_0}(x,y;0)=\delta(x-y) \nonumber \\
&&\left(\frac{\partial}{\partial\beta}+K\right)K_{K}(x,y;\beta)=0
\qquad , \qquad K_{K}(x,y;0)=\delta(x-y) \label{heke} \qquad ,
\end{eqnarray}
which provide an alternative way of writing the $K_0$- and $K$-heat
traces:
\[
{\rm Tr}_{L^2}e^{-\beta
K_0}=\lim_{l\to\infty}\int_{-\frac{l}{2}}^\frac{l}{2}\, dx \,
K_{K_0}(x,x;\beta) \qquad , \qquad {\rm Tr}_{L^2}e^{-\beta
K}=\lim_{l\to\infty}\int_{-\frac{l}{2}}^\frac{l}{2}\, dx \,
K_{K}(x,x;\beta) \qquad .
\]
Note that the form of the $K$-heat equation (\ref{heke}), $
\left(\frac{\partial}{\partial\beta}+K_0-U(x)\right)K_K(x,y;\beta)=0$,
suggests a solution based on the $K_0$-heat kernel: $
K_K(x,y;\beta=C_K(x,y;\beta)K_{K_0}(x,y;\beta)$. The density
$C_K(x,y;\beta)$ satisfies the infinite temperature condition
$C_K(x,y;0)={\mathbb I}_{N\times N}$ and the transfer equation:
\begin{equation}
\left(\frac{\partial}{\partial\beta} +\frac{x-y}{\beta}
\frac{\partial}{\partial x}-\frac{\partial^2}{\partial
x^2}\right)C_{K}(x,y;\beta)=U(x)C_{K}(x,y;\beta) \label{trans} \quad
.
\end{equation}
Next we seek a power series solution,
$C_{K}(x,y;\beta)=\sum_{n=0}^\infty \, c_n(x,y)\beta^n$, of
(\ref{trans}), which becomes tantamount to the recurrence relations:
\begin{equation}
n c_n(x,y)+(x-y)\frac{\partial c_n}{\partial
x}(x,y)=\frac{\partial^2 c_{n-1}}{\partial x^2}(x,y)+
U(x)c_{n-1}(x,y) \label{recr} \quad .
\end{equation}
In fact, only the densities at coincident points $x=y$ on the line
are needed. We introduce the notation ${}^{(k)}C_n(x)=\lim_{x\to
y}\frac{\partial^k c_n}{\partial x^k}(x,y)$ to write the recurrence
relations for the Seeley densities (and their derivatives) in the
abbreviated form:
\[
{}^{(k)} C_n(x) =\frac{1}{n+k} \left[ {}^{(k+2)} C_{n-1}(x) -
\sum_{j=0}^k {k \choose j} \frac{\partial^j U(x)}{\partial x^j}
{}^{(k-j)} C_{n-1}(x) \right] \qquad .
\]
The (Seeley) coefficients $c_n(K)$ are the integrals over the
infinite line of the densities $c_n(x,x)$, i.e.,
$c_n(K)=\int_{-\infty}^\infty dx c_n(x,x)$.

\end{document}